\documentclass{jltp}
\usepackage{graphicx}
\title{Single-electron transport driven by surface acoustic waves: moving
quantum dots versus short barriers}

\author{P.~Utko,$^1$ J.~Bindslev Hansen,$^2$ P.~E.~Lindelof,$^1$
   C.~B.~S{\o}rensen,$^1$ \\and K.~Gloos$^3$}

\address{
  $^1$Nano-Science Center and Niels Bohr Institute, University of Copenhagen,\\
   Universitetsparken 5, DK-2100 Copenhagen, Denmark\\
  $^2$Department of Physics, Technical University of Denmark,\\
  DK-2800 Lyngby, Denmark\\
  $^3$Wihuri Physical Laboratory, Department of Physics, University of Turku,\\
     FIN-20014 Turku, Finland}

\runninghead{P.~Utko {\it et al.}}{Single-electron transport driven by surface
acoustic waves}
\begin{document}
\maketitle
\date{(Received: March 24, 2006)}%

\begin{abstract}
We have investigated the response of the acoustoelectric current
driven by a surface-acoustic wave through a quantum point contact in
the closed-channel regime. Under proper conditions, the current
develops plateaus at integer multiples of $ef$ when the frequency
$f$ of the surface-acoustic wave or the gate voltage $V_g$ of the
point contact is varied. A pronounced 1.1\,MHz beat period of the
current indicates that the interference of the surface-acoustic wave
with reflected waves matters. This is supported by the results
obtained after a second independent beam of surface-acoustic wave
was added, traveling in opposite direction. We have found that two
sub-intervals can be distinguished within the 1.1\,MHz modulation
period, where two different sets of plateaus dominate the
acoustoelectric-current versus gate-voltage characteristics. In some
cases, both types of quantized steps appeared simultaneously, though
at different current values, as if they were superposed on each
other. Their presence could result from two independent quantization
mechanisms for the acoustoelectric current. We point out that short
potential barriers determining the properties of our nominally long
constrictions could lead to an additional quantization mechanism,
independent from those described in the standard model of 'moving
quantum dots'.


\end{abstract}



\section{INTRODUCTION}
\label{intro} Single-electron transport through a quantum-point
contact (QPC), driven by a surface-acoustic wave (SAW), is
considered to be an attractive method towards a quantum standard of
electrical current.\cite{Shilton1996,Talyanskii1997} Such a standard
would close the quantum-metrological triangle of electrical units of
current, voltage, and resistance, and allow us to determine the
electron charge $e$ and the Planck constant
$h$.\cite{Zimmermann2003,Flowers2001} The SAW-based approach offers
much higher operational frequencies (several 1\,GHz) and output
currents ($\sim1\,$nA) than other types of electron
pumps,\cite{Kouwenhoven1991,Keller1999,Switkes1999,Pekola1994} which
are limited to frequencies of order of 10\,MHz and output currents
of about 10\,pA. However, the low accuracy\cite{Cunningham1999,Cunningham2000,Cunningham2000-77,%
Ebbecke2000,Ebbecke2002,Utko2003,Gloos2004} of the SAW pumps
($\sim100\,$ppm) still prevents their metrological applications.

The SAW-driven single-electron devices have also been considered for
other purposes. Foden {\it et al.}\cite{Foden2000} suggested to use
them as a key component of a single-photon source. In such a scheme,
the SAW propagates along a lateral n-i-p junction. Its dynamic
potential captures electrons in the n-type region, from the
two-dimensional electron gas (2DEG) reservoir at the entrance to a
QPC. Single electrons per SAW cycle are then transferred across the
point contact and injected into the p-type region. They recombine
there with holes, emitting a single photon per injected electron. In
another proposal, Barnes {\it et al.}\cite{Barnes2000,Furuta2004}
suggested to use spins of the single electrons trapped in the SAW
minima as qubits for quantum-computation applications. They
demonstrated theoretically the feasibility of such an approach for
one- and two-qubit operations. This would involve a proper design of
adjacent channels for the SAW-driven electrons and a pattern of
magnetic and non-magnetic surface gates.

Qualitatively, the SAW pumps operate as follows. An interdigital
transducer (IDT) generates - via the piezoelectric effect - elastic
waves on the surface of a gallium arsenide sample that contains a
2DEG beneath its surface. Due to piezoelectricity, the mechanical
component of the SAW is accompanied by an electrostatic potential
which can induce a current flow across a closed quantum point
contact. Shilton {\it et al.}\cite{Shilton1996,Talyanskii1997}
suggested that electrons are trapped in the 'moving quantum dots',
local minima of the dynamic SAW potential that travel at the sound
velocity up the potential hill of the QPC. The occupancy of a
specific dot is determined by the Coulomb interaction between the
electron population. Thus, an integer number $n$ of electrons is
transported per SAW cycle across the constriction.

Different aspects of this model were later discussed in a number of
experimental\cite{Cunningham1999,Cunningham2000,Cunningham2000-77,%
Ebbecke2000,Ebbecke2002,Utko2003,Gloos2004} and
theoretical\cite{Aizin1998,Flensberg1999,Robinson2001,Kashcheyevs2004}
studies. Deviations from perfectly flat plateaus in the SAW-driven
acoustoelectric (AE) current were attributed to electron tunneling
or thermal activation, either into\cite{Flensberg1999} or out
of\cite{Robinson2001} the moving quantum dot. Recently, Fletcher
{\it et al.} suggested an alternative mechanism involving a static
quantum dot, either impurity-induced\cite{Fletcher2003} or
fabricated on purpose.\cite{Ebbecke2004} The potential barriers of
such a static dot are tuned like a turnstile by the low-power SAW.

This short overview already indicates that different effects could
be involved in the SAW-driven single electron transport, and that
essential transport mechanisms have not been clearly identified yet.
Different mechanisms might dominate different samples. Neither can
we rule out possible transitions between different mechanisms in the
same device.

Cunningham {\it et al.}\cite{Cunningham2000,Cunningham2000-77}
demonstrated that the precision of the AE current quantization can
be improved by using shallow-etched QPCs. However, till today the
absolute accuracy of the AE current is not sufficient for
practical applications as a metrological standard.\cite{Cunningham1999,%
Cunningham2000,Cunningham2000-77,Ebbecke2000,Ebbecke2002,Utko2003,Gloos2004}
Current plateaus, as a function of gate voltage $V_g$ or SAW
frequency $f$, are not really flat, leading to a rather arbitrary
selection of their exact value. Even at nearly optimum conditions,
the AE current at such plateaus is smaller than the ideally expected
multiples of $ef$. This reduction as well as the finite slope of the
plateaus could result from the enhanced temperature of the 2DEG due
to the applied rf power. On the other hand, this could also result
from the electron tunneling through the walls of the moving quantum
dots which always have a finite thickness and depth. These topics
might distract attention from others which are less obvious. Here,
we discuss some of the more basic properties or our SAW devices,
their response to the SAW frequency (phase) and the gate voltage.

\section{EXPERIMENTAL DETAILS}
\label{exp}
\begin{figure}
 \begin{center}
  \includegraphics[width=9cm]{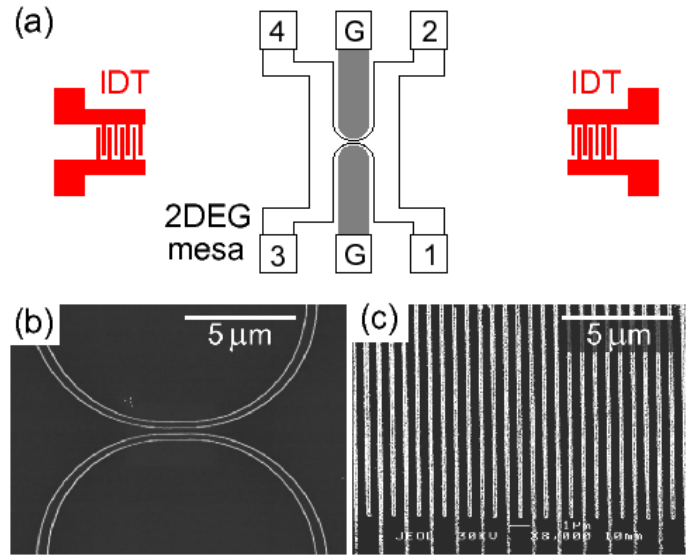}
 \end{center}
\caption{(Color on-line) (a) Schematic diagram of sample layout: A
quantum point contact (QPC) is defined in the 2DEG mesa using a
shallow-etch technique. Four Ohmic contacts (1-4) provide electrical
connection to the electron reservoirs on both sides of the
constriction. The shaded regions are the reservoirs which serve as
side gates (G). Two interdigital transducers (IDTs) are deposited on
opposite sides of the mesa. Scanning-electron micrographs show (b)
the QPC and (c) part of the IDT. The two pairs of bright lines in
(b) mark the edges of the shallow-etched trenches that separate the
side gates (top and bottom) from the 2DEG (left to right). The
parallel white lines in (c) are the IDT fingers.}
\label{setup}
\end{figure}

Figure \ref{setup} shows the layout of our devices: Two
aluminum interdigital transducers (IDTs) could be used to generate
the SAW. They were deposited 2.6\,mm apart, on both sides of a 2DEG
mesa with a QPC in the center. The IDT electrode spacing set the
fundamental acoustic wavelength of the transducers to about
$1.15\,\mu$m and their center frequency to around 2.45\,GHz. The
GaAs/AlGaAs heterostructure had a mobility of $105\,$m$^2/$(Vs) and
a carrier density of $2.8 \cdot 10^{15}$\,m$^{-2}$, measured in the
dark at 10\,K. The QPC was patterned by electron-beam lithography.
Two semicircular shallow-etched trenches formed a smooth
constriction between the two electron reservoirs, whereas the large
areas of the 2DEG across the channel served as side gates. The
trenches had a curvature radius of 5.0, 7.5 or 10.0\,$\mu$m. They
were 200\,nm wide and 40\,nm deep.

The rf excitation of power $P$, from an Agilent 8648D or a
Hewlett-Packard HP8673B microwave generator, could be applied to one
of the two IDTs, or split up and simultaneously fed to both
transducers. With a phase shifter and an attenuator in one of the rf
lines, the relative magnitude and the relative phase of both signals
could then be varied and adjusted. A low-noise current preamplifier
detected the acoustoelectric current.

The samples were investigated either in a $^3$He refrigerator
operating at 1.2\,K or a $^4$He refrigerator operating at 1.8\,K.
The lower base temperature of the $^3$He cryostat did not help to
improve the results. In both systems, the 2DEG of our devices was
heated up to around 5\,K at the typically applied rf powers of
around $10-15\,$dBm.\cite{Utko2006}

\subsection{Finding the quantized steps}

Finding the AE plateaus in the $I(V_g)$ characteristics was not
always an easy task due to a large number of parameters that had to
be tuned: the gate and bias voltage, the SAW power and frequency.
For the double-beam configuration, this also included the phase and
the relative magnitude of the two counter-propagating SAW beams.
Usually, a large number of scans within this parameter space was
necessary to find quantized steps in the AE current or to discard
the device.

\begin{figure}
 \begin{center}
  \includegraphics[width=11cm]{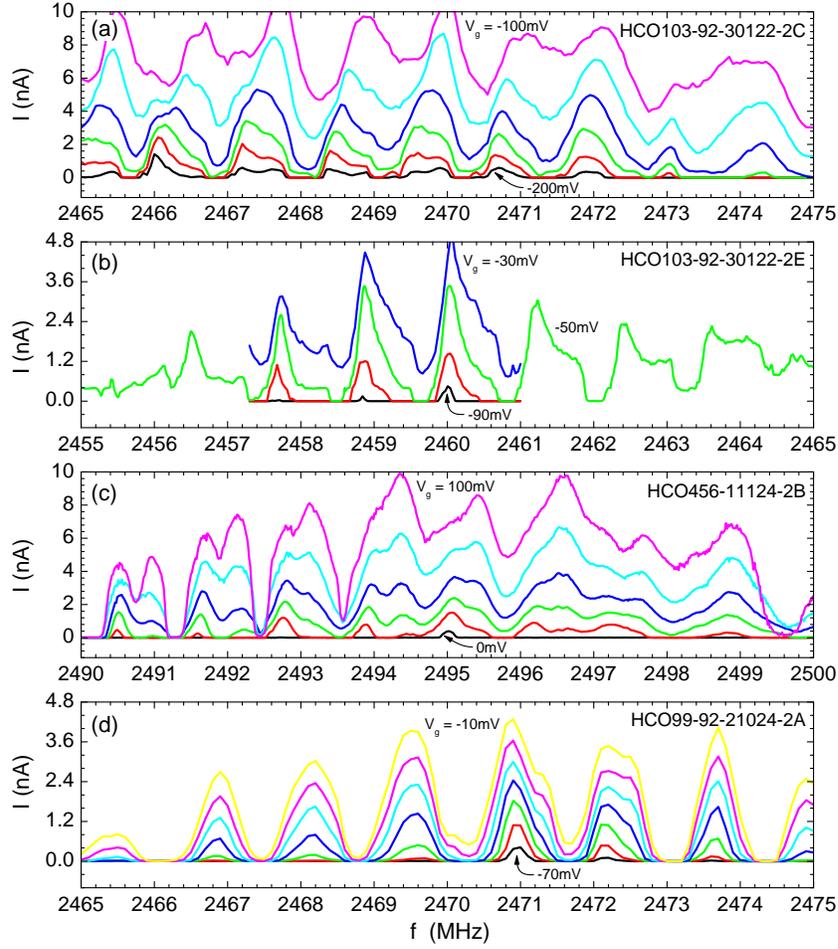}
 \end{center}
\caption{(Color on-line) Acoustoelectric current $I$ versus SAW
frequency $f$ in the indicated range of gate voltages $V_g$ below
conductance pinch-off. The measurements were performed at $T =
1.8\,$K for four different samples: (a) HCO103-92-30122-2C, (b)
HCO103-92-30122-2E, (c) HCO456-11124-2B, and (d) HCO99-92-21024-2A.}
\label{frequency}
\end{figure}

\begin{figure}
 \begin{center}
  \includegraphics[width=11cm]{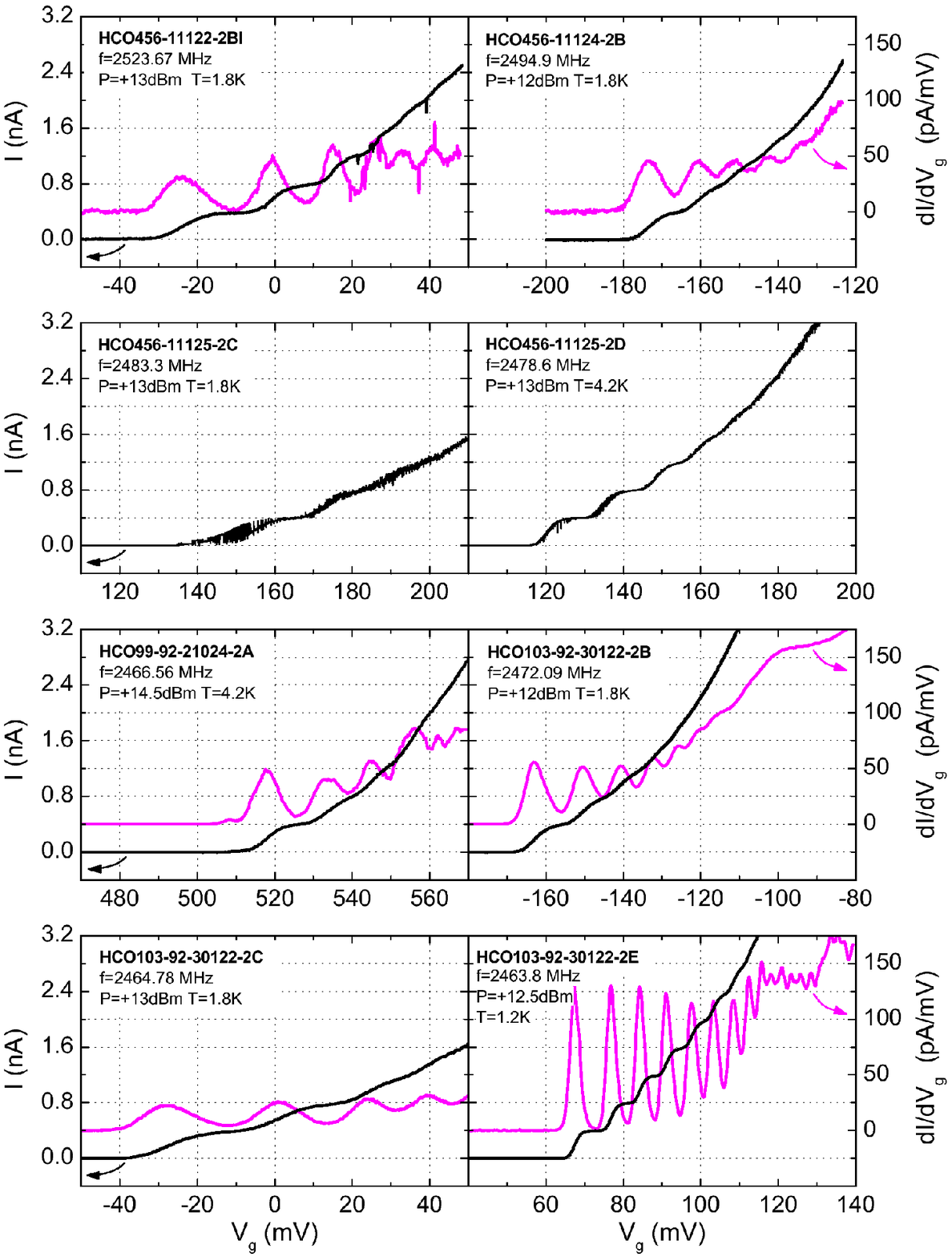}
 \end{center}
\caption{(Color on-line) (dark) Acoustoelectric current $I$ and
(light) transconductance $dI/dV_g$ as a function of gate voltage
$V_g$ for all eight working SAW devices. Frequency $f$, rf power
$P$, as well as the base temperature $T$ of the refrigerator are
indicated in the consecutive panels. In some of the curves,
pronounced random telegraph noise is present.} \label{typical}
\end{figure}

The main components of the device, the IDTs and the QPC, were first
characterized by measuring the transmittance of the SAW delay line
and the conductance $G$ of the QPC. The $G(V_g)$ characteristics
provided information on the pinch-off gate voltage for conductance,
whereas the transmittance indicated the pass band at which the SAW
transducers should be operated.

The typical optimization procedure was then as follows. The gate
voltage on the QPC was set below the conductance pinch-off
(typically $\sim 100\,$mV below) and an rf power of around
$10-15\,$dBm was applied to one of the IDTs. The SAW frequency was
scanned to determine the efficiency of each transducer in terms of
the AE current carried across the QPC. Figure~\ref{frequency} shows
typical $I(f)$ curves of that type. Note the periodic $1.1\,$MHz
oscillations resulting from the SAW reflections from the second,
unconnected IDT.\cite{Talyanskii1997} Those oscillations,
characteristic for all our devices, are discussed in the following
sections, along with their finer features like the two shifted
1.1\,MHz beats clearly visible in Fig.~\ref{frequency}(c).

The $I(f)$ traces were inspected for any indications of quantized
steps, a flattening of $I(f)$ near integer multiples of $ef$.
Intervals of roughly 1\,MHz were thus selected around current peaks
with such anomalies. The $I(V_g)$ characteristics, similar to those
in Fig. \ref{typical}, were then recorded at fixed frequencies from
those ranges, in steps of $\Delta f \approx 0.1\,$MHz. To enhance
the resolution of the current plateaus, we monitored as well the
transconductance $dI/dV_g$. This was done by adding a small 117\,Hz
modulation of $dV_g \approx 0.5\,$mV to the gate voltage, and
measuring the resulting ac component $dI$ of the acoustoelectric
current with a lock-in amplifier. With those precise $I(V_g ,f)$ and
$dI(V_g ,f)/dV_g$ maps, the AE transitions could be localized, and
extensive scans over the interesting parameter range could be
initiated.

\subsection{Counter-propagating SAW beams}

The acoustoelectric current carried across the QPC relies on the
dynamic modulation of the electrostatic potential imposed by the
SAW. This modulation can be seriously affected in the presence of
other spurious signals. For example, the main SAW beam could
interfere with an additional counter-propagating beam, which results
from the SAW reflections from the second (unconnected)
transducer.\cite{Talyanskii1997,Datta1986} In fact, we consider such
as scenario as the most likely explanation for strong current
oscillations with respect to frequency, typical for all our devices
(Fig.~\ref{frequency}).

Two counter-propagating SAW beams can be described as a
superposition of a traveling and a (usually weaker) standing wave.
The position of the standing wave nodes and antinodes with respect
to the QPC is then determined by the phase difference between the
two beams. Assuming that the backward beam originates from the SAW
reflections from the second transducer, the phase shift depends on a
double distance $2L$ between the point contact and the unconnected
IDT. In our structures $2L \approx 2.6\,$mm which should result in a
frequency modulation of $\Delta f = v_{SAW}/2L \approx 1.1\,$MHz,
remarkably close to the beat period observed in our measurements.
Here, $v_{SAW} \approx 2800\,$m/s is the velocity of the surface
wave. In a previous study by Talyanskii \emph{et
al.},\cite{Talyanskii1997} a larger separation between the QPC and
the second transducer $2L \approx 4\,$mm yielded a correspondingly
smaller beat period of $\Delta f \approx 0.7\,$MHz. In addition, no
current oscillations were observed there for a sample with only one
IDT.

We note that the SAW potential might also interfere (cross-talk)
with the electromagnetic wave irradiated by the transducer. Since
the airborne signal reaches the QPC almost instantaneously, the
resulting frequency modulation in our devices should then be around
$\Delta f = v_{SAW}/L \approx 2.2\,$MHz, unlike the typically
observed $1.1\,$MHz beat.

The second, counter-propagating SAW beam can also be generated on
purpose by applying a microwave signal to both
transducers.\cite{Cunningham1999,Utko2003} This allows a direct
control over the relative magnitude and the relative phase between
the two beams. Experiments on the SAW-driven single-electron pumps
in a double-beam configuration revealed notable changes in the
$I(V_g)$ characteristics in response to phase
variations.\cite{Cunningham1999} In particular, the slope of the
first AE plateau in $I(V_g)$ was significantly reduced for some
phase settings. This already indicated that the second SAW beam
could be used to further improve the accuracy of the AE current
quantization, if only the phase difference between the two
counter-propagating beams was properly
optimized.\cite{Cunningham1999,Utko2003}

We emphasize the fact that two facing transducers were deposited on
opposite sides of the QPC for all our samples, even though a single
IDT is sufficient for the SAW pump operation. This was done in order
to increase the yield of working devices and to enable experiments
in a double-beam configuration. However, this also implied that no
direct comparison was possible with devices comprising only one
transducer for which the backward beam should be effectively
dampened.

\section{EFFECT OF THE SAW FREQUENCY}

Our measurements of the AE current $I$ and the transconductance
$dI/dV_g$, versus both the gate voltage $V_g$ and the SAW frequency
$f$, clearly show the current modulation with respect to frequency,
with a period of about 1.1\,MHz. However, within this period two
frequency intervals can be distinguished for which apparently
different sets of AE plateaus dominate the $I(V_g)$ characteristics.
This feature shows up whether the SAW is generated by one IDT only
(single-beam configuration) or by two facing IDTs (double-beam
configuration). We point out that corresponding intervals can also
be resolved when the relative phase between the two
counter-propagating SAW beams is varied instead of frequency. Two of
our devices are studied in detail: 2C (= HCO103-92-30122-2C) and 2E
(= HCO103-92-30122-2E).

\begin{figure}
\includegraphics[width=1\textwidth]{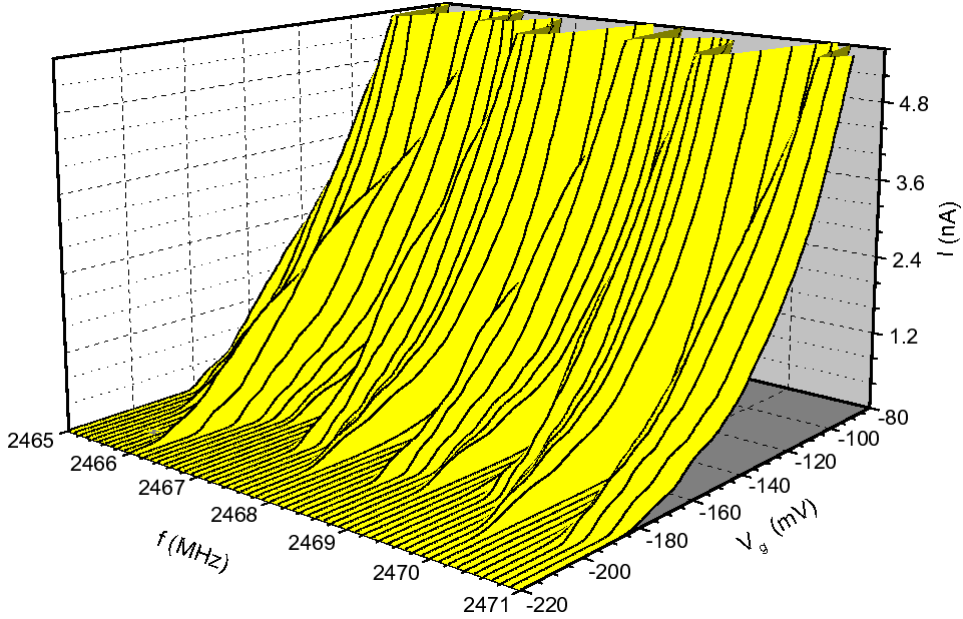}
\caption{(Color on-line) AE current $I$ versus gate voltage $V_g$
and SAW frequency $f$.  The measurement was performed in a
single-beam configuration. Device 2C at $P = +9.8\,$dBm and $T =
1.8\,$K.} \label{3dim-2C}
\end{figure}

\begin{figure}
\includegraphics[width=1\textwidth]{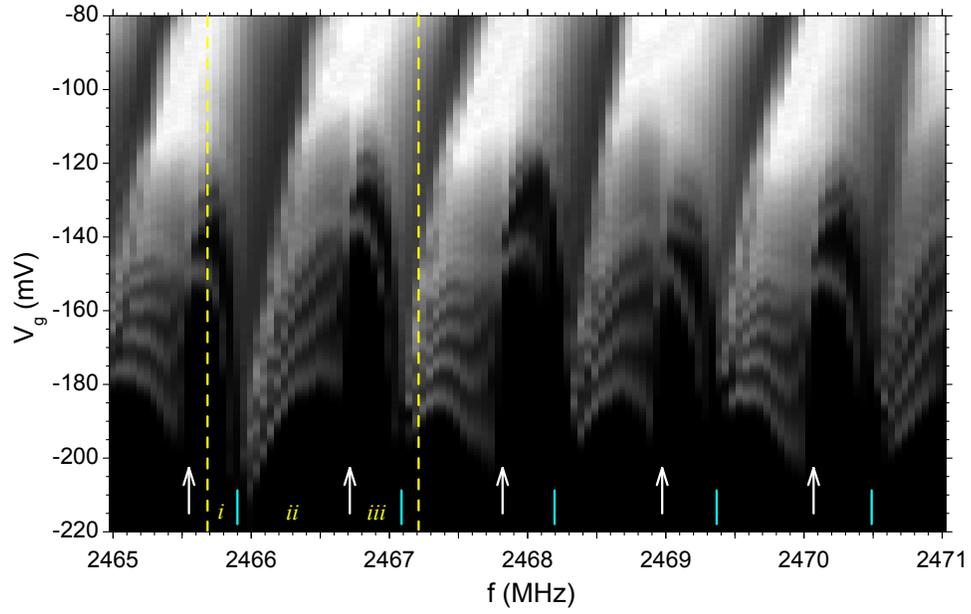}
\caption{(Color on-line) Gray-scale coded plot of transconductance
$dI/dV_g$ with respect to SAW frequency $f$ and gate voltage $V_g$.
Dark (light) indicates small (large) values of the current
derivative.  Note the periodic structure with a modulation period of
about 1.1\,MHz. In addition, two smaller intervals can be
distinguished within such a period where different sets of AE
plateaus seem to prevail. The arrows and bars mark the frequencies
around which one set is replaced by another one, see text for
details. The dashed lines indicate the frequency range from which
the traces in Fig.~\ref{results-2C} were selected. Device 2C at $P =
+9.8\,$dBm and $T = 1.8\,$K.} \label{greyscale-2C}
\end{figure}

\begin{figure}
  \begin{center}
   \includegraphics[width=1\textwidth]{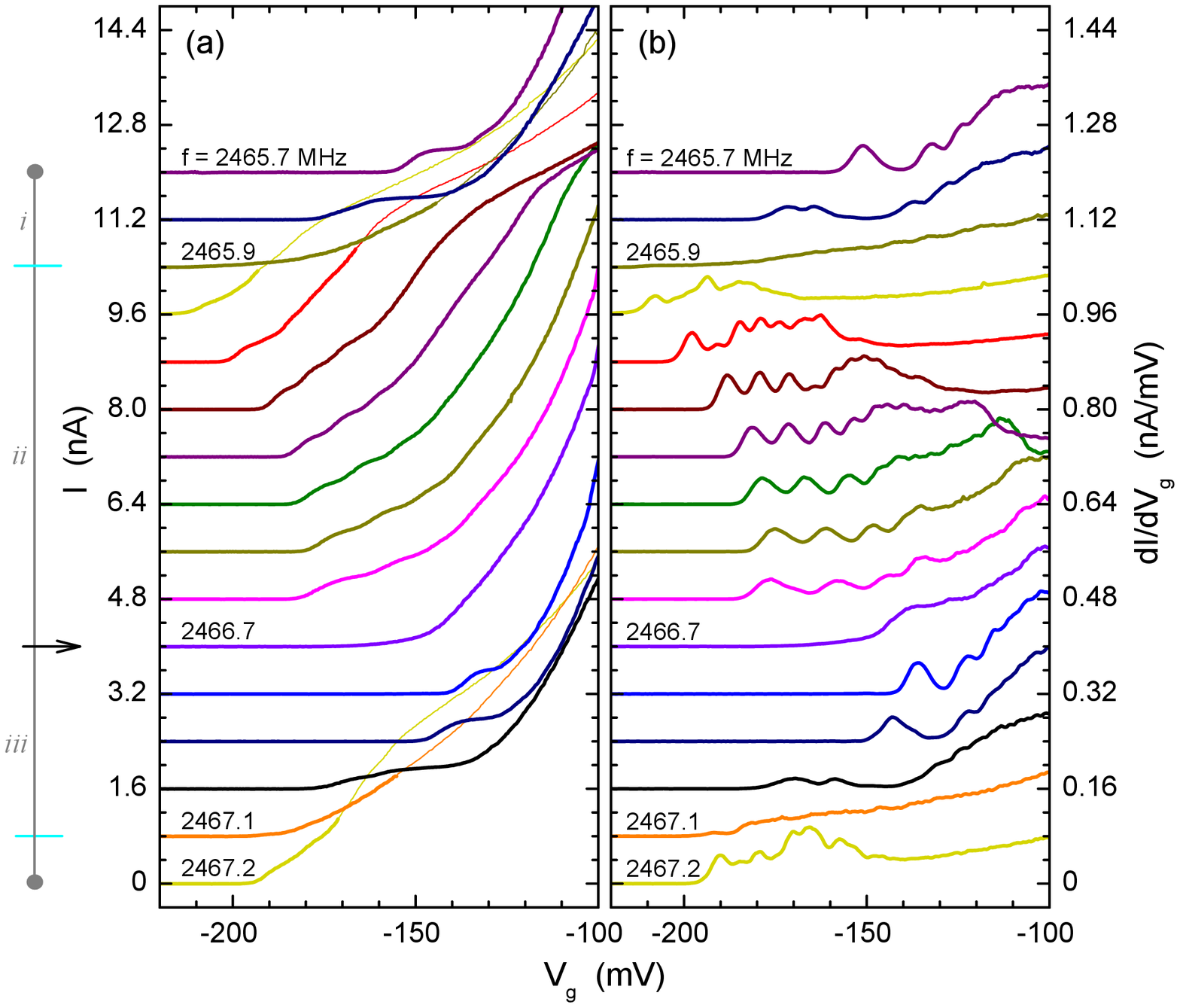}
  \end{center}
\caption{(Color on-line) (a) AE current $I$ and (b) transconductance
$dI/dV_g$ as a function of gate voltage $V_g$. The 16 current
(transconductance) traces were recorded at fixed SAW frequencies
from 2465.7 to 2467.2\,MHz, in steps of 0.1\,MHz. The curves were
successively offset in the vertical direction by 0.8\,nA
(0.08\,nA/mV). Device 2C at $P = +9.8\,$dBm, and $T = 1.8\,$K.}
\label{results-2C}
\end{figure}

\subsection{Single-beam configuration}
\label{single-beam}

Figure~\ref{3dim-2C} shows the acoustoelectric current with respect
to both the gate voltage and the SAW frequency for sample 2C. During
the measurement, only one IDT was used to generate the SAW
(single-beam configuration). The AE plateaus could be resolved for
all frequencies within the shown 6\,MHz interval. However, their
slope as well as their extension along the gate-voltage axis
responded very sensitively to variations in frequency.

The AE transitions are easier to resolve in the current derivative.
Figure~\ref{greyscale-2C} shows a gray-scale plot of the
transconductance $dI/dV_g$, revealing a periodic structure with
respect to frequency with a beat period of roughly $1.1\,$MHz.
However, transconductance minima corresponding to current plateaus
do not evolve smoothly over the entire 1.1\,MHz interval. On
increasing the SAW frequency, broad plateaus in $I(V_g)$ are
abruptly replaced by densely-packed quantized steps. Frequencies
around which such transitions occur are marked with bars. Arrows
indicate another type of transition where the AE current drastically
shifts its onset along the $V_g$-axis in response to a small change
in frequency. This displacement can be as large as $\Delta V_g
\approx 50\,$mV over $\Delta f \approx 0.2\,$MHz, as for example
around $f \approx 2467.8\,$MHz.

Those features become more apparent in Fig.~\ref{results-2C} which
shows the selected $I(V_g)$ and $dI(V_g)/dV_g$ characteristics from
Figs.~\ref{3dim-2C} and \ref{greyscale-2C}, respectively. In the
following, we describe the frequency evolution of those curves in
more detail. As in Fig.~\ref{greyscale-2C}, the bars and arrows mark
the traces around which the discussed transitions occur.

(\emph{i}) When the SAW frequency is incremented from 2465.7\,MHz,
the first plateau in $I(V_g)$ broadens with respect to the gate
voltage. At the same time, the current onset shifts towards smaller
values of $V_g$. At $f = 2465.9\,$MHz, the first quantized step can
no longer be resolved in either the AE current or the
transconductance.

\begin{figure}
\includegraphics[width=1\textwidth]{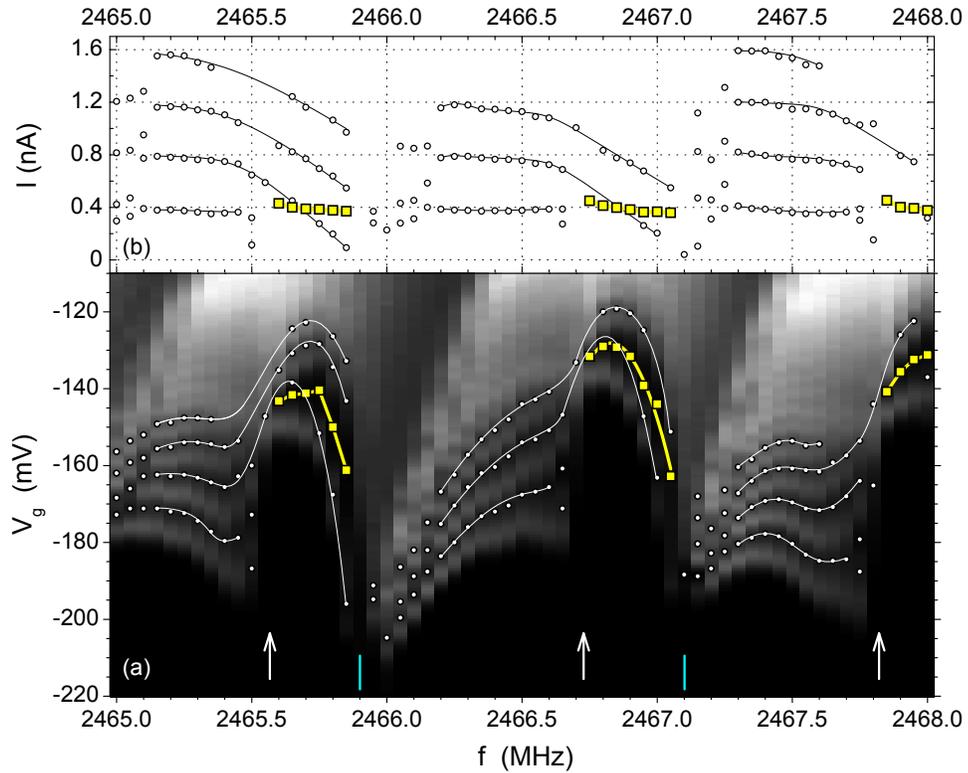}
\caption{(Color on-line) (a) Gray-scale plot of transconductance
$dI/dV_g$ with respect to SAW frequency $f$ and gate voltage $V_g$.
Dark (light) indicates small (large) values of the derivative. Open
circles and full squares mark the position of transconductance
minima. (b) AE current $I$ at the minima indicated in (a). In both
(a) and (b), two different kinds of AE plateaus can be
distinguished. Solid lines through the data points are guides to the
eye. Device 2C at $P = +9.8\,$dBm and $T = 1.8\,$K.}
\label{plateaus-2C}
\end{figure}

(\emph{ii}) When the frequency is increased above $2465.9\,$MHz, the
AE plateaus reappear in the $I(V_g)$ traces. However, they are much
less pronounced than those in (\emph{i}), as both their slope and
number per gate-voltage interval has increased. This new set of
plateaus evolves smoothly until $f \approx 2466.7\,$MHz, where the
onset of the $I(V_g)$ characteristics abruptly shifts towards higher
gate voltages. For the traces obtained at $f = 2466.6$ and
2466.8\,MHz, this displacement is as large as $\Delta V_g \approx
40\,$mV.

(\emph{iii}) Above 2466.7\,MHz, another set of quantized steps
develops in $I(V_g)$. It continues until $f = 2467.1\,$MHz, behaving
in a similar way as the corresponding set in (\emph{i}).

In both Figs.~\ref{greyscale-2C} and ~\ref{results-2C}, two
frequency intervals can be distinguished within a beat period of
$\sim1.1\,$MHz where two different sets of AE plateaus dominate the
$I(V_g)$ characteristics. Those two sets seem to replace each other
around certain frequencies, indicated by the arrows and bars.
However, in some cases, both sets can also appear simultaneously in
the $I(V_g)$ traces, as if they were superposed onto each other. The
dominating set consists then of broad current plateaus that are
well-defined at the expected multiples of $ef$. On the other hand,
weakly-pronounced steps belonging to the second set are formed below
those ideal values, as shown in Fig.~\ref{plateaus-2C}.

Both sets respond differently to the SAW frequency as becomes
apparent, for example, in the following range $f = 2465.7 -
2465.9\,$MHz in Fig.~\ref{plateaus-2C}. When the frequency is
incremented within such interval, one plateau in $I(V_g)$ remains
close to the expected value of $ef \approx 400\,$pA, while the other
set of quantized steps appears at lower and lower currents. The
plateau at $I = ef$ is broad and well-pronounced with respect to the
gate voltage, while the steps away from the expected multiples of
$ef$ are barely indicated in the $I(V_g)$ characteristics.

\subsection{Double-beam configuration}
\label{double-beam}

Two separate sets of current plateaus, appearing within the 1.1\,MHz
beat period, could also be observed when the device was operated in
a double-beam configuration.

\begin{figure}
\includegraphics[width=1\textwidth]{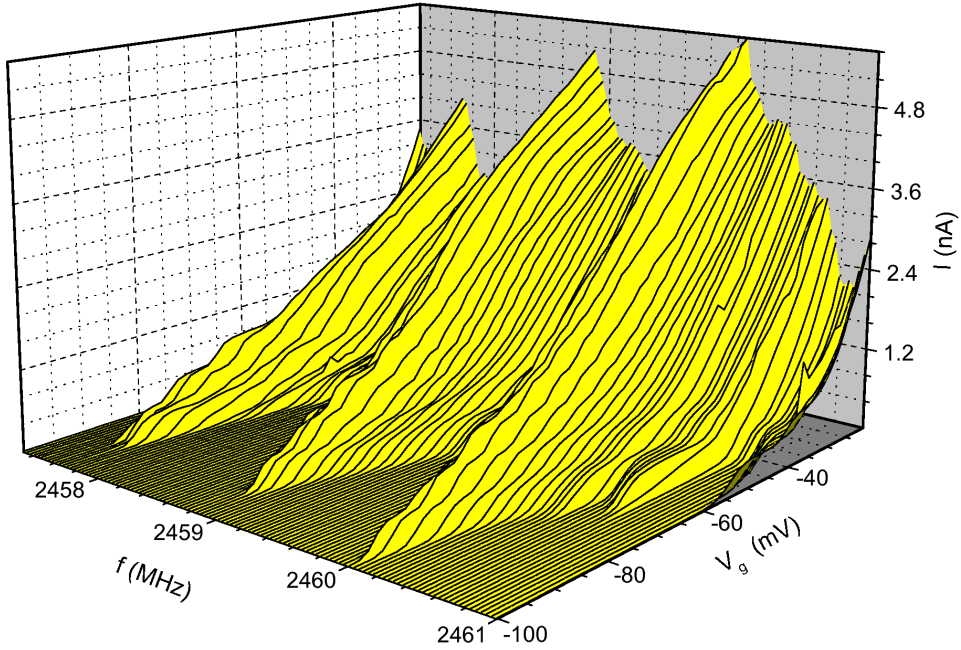}
\caption{(Color on-line) AE current $I$ versus gate voltage $V_g$
and SAW frequency $f$. The measurement was performed in a
double-beam configuration. The microwave signal of $P = +13.5\,$dBm
was applied to the driving transducer, while the excitation applied
to the second IDT was attenuated by 8\,dB. Device 2E at $T =
1.2\,$K.} \label{3dim-2E}
\end{figure}

\begin{figure}
\includegraphics[width=1\textwidth]{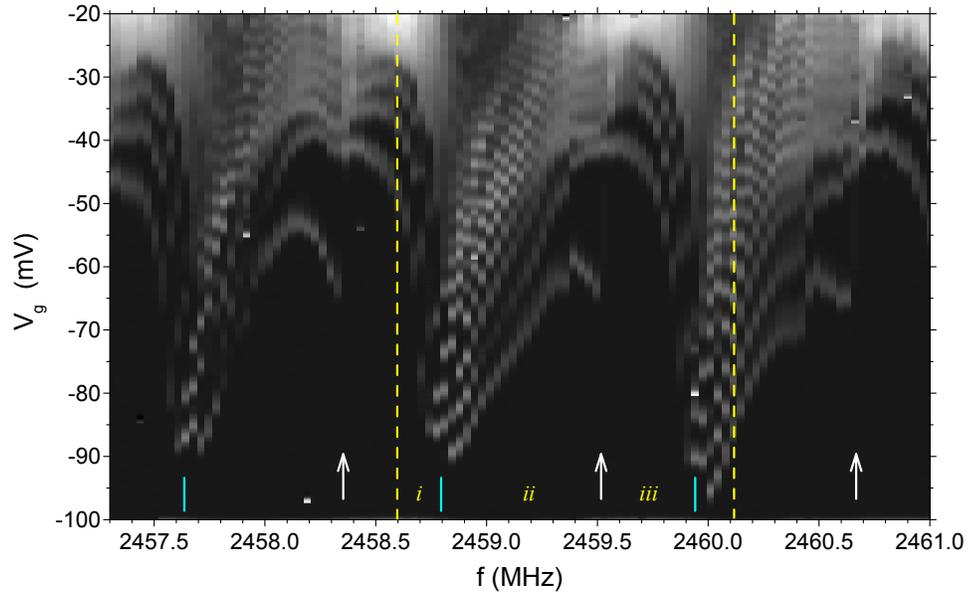}
\caption{(Color on-line) Gray-scale plot of transconductance
$dI/dV_g$ with respect to SAW frequency $f$ and gate voltage $V_g$.
Dark (light) indicates small (large) current derivative. The
measurement was performed in a double-beam configuration. Bars and
arrows indicate frequencies around which one set of AE plateaus is
replaced by another. The dashed lines indicate the frequency range
from which the traces in Fig.~\ref{results-2E} were selected. Device
2E at $T = 1.2\,$K.} \label{greyscale-2E}
\end{figure}

Figures~\ref{3dim-2E} and \ref{greyscale-2E} show the
acoustoelectric current and the transconductance, respectively, for
device 2E. During the measurement, two counter-propagating SAW beams
were generated on purpose from the two transducers on opposite sides
of the QPC. An rf power of $P = 13.5\,$dBm was applied to the
driving IDT, while the excitation applied to the second transducer
was attenuated by 8\,dB. In order to obtain the flattest plateaus in
the acoustoelectric current, the relative phase between the two
counter-propagating SAW beams was optimized. The slope of the first
plateau was thus reduced by an order of magnitude with respect to
the case when the second beam was off.\cite{Utko2003}

As for the single-beam configuration, both the AE current
(Fig.~\ref{3dim-2E}) and the transconductance
(Fig.~\ref{greyscale-2E}) reveal a very clear 1.1\,MHz-beating. Two
frequency intervals can be distinguished within this period, where
two different sets of the AE plateaus dominate the $I(V_g)$
characteristics.

\begin{figure}
 \begin{center}
  \includegraphics[width=1\textwidth]{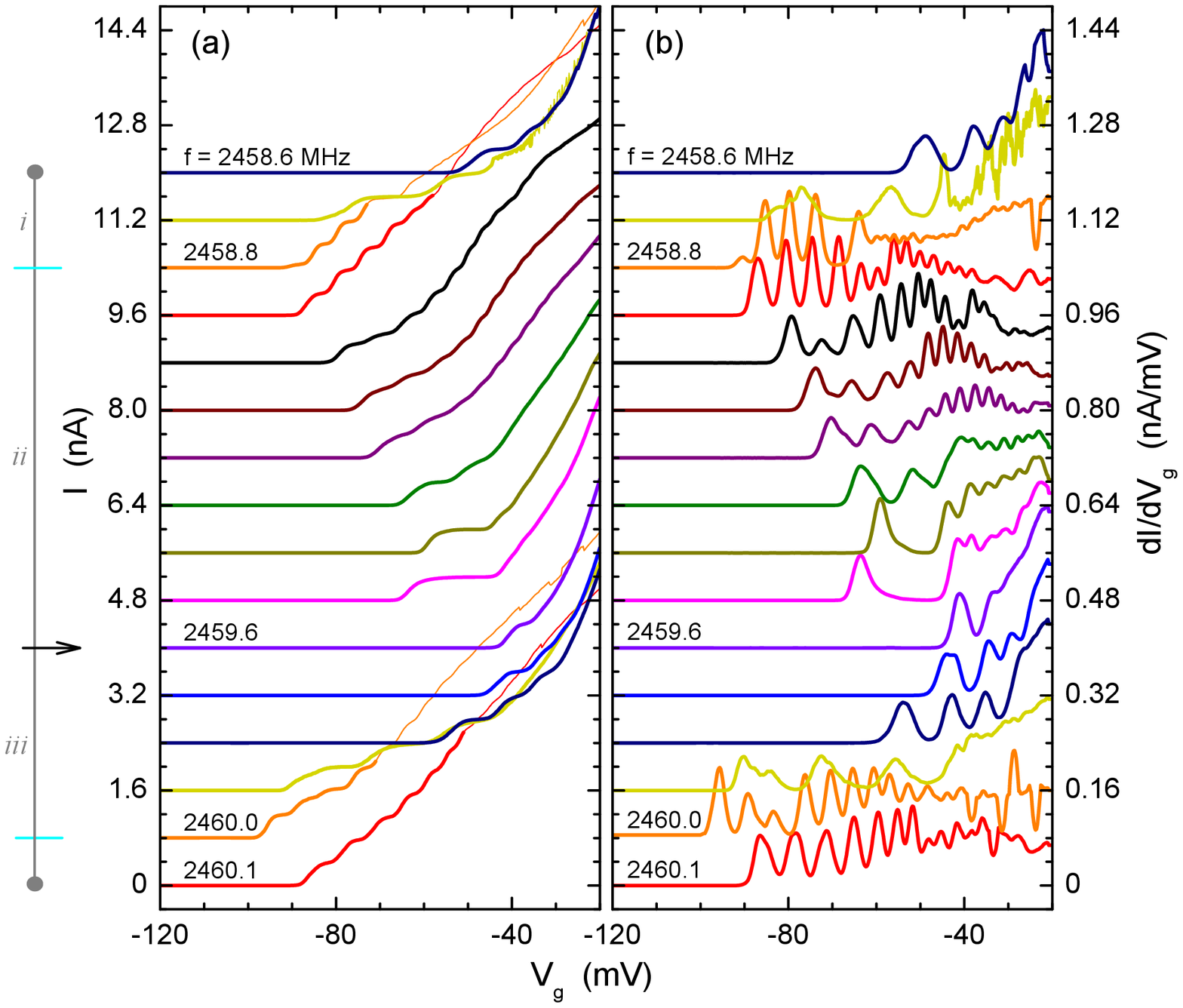}
 \end{center}
\caption{(Color on-line) (a) AE current $I$ and (b) transconductance
$dI/dV_g$ as a function of gate voltage $V_g$. The 16 current
(transconductance) traces were recorded at fixed SAW frequencies
from 2458.6 to 2460.1\,MHz, in steps of 0.1\,MHz. The curves were
successively offset in the vertical direction by 0.8\,nA
(0.08\,nA/mV). Device 2E at $T = 1.2\,$K.} \label{results-2E}
\end{figure}
\begin{figure}
 \begin{center}
  \includegraphics[width=1\textwidth]{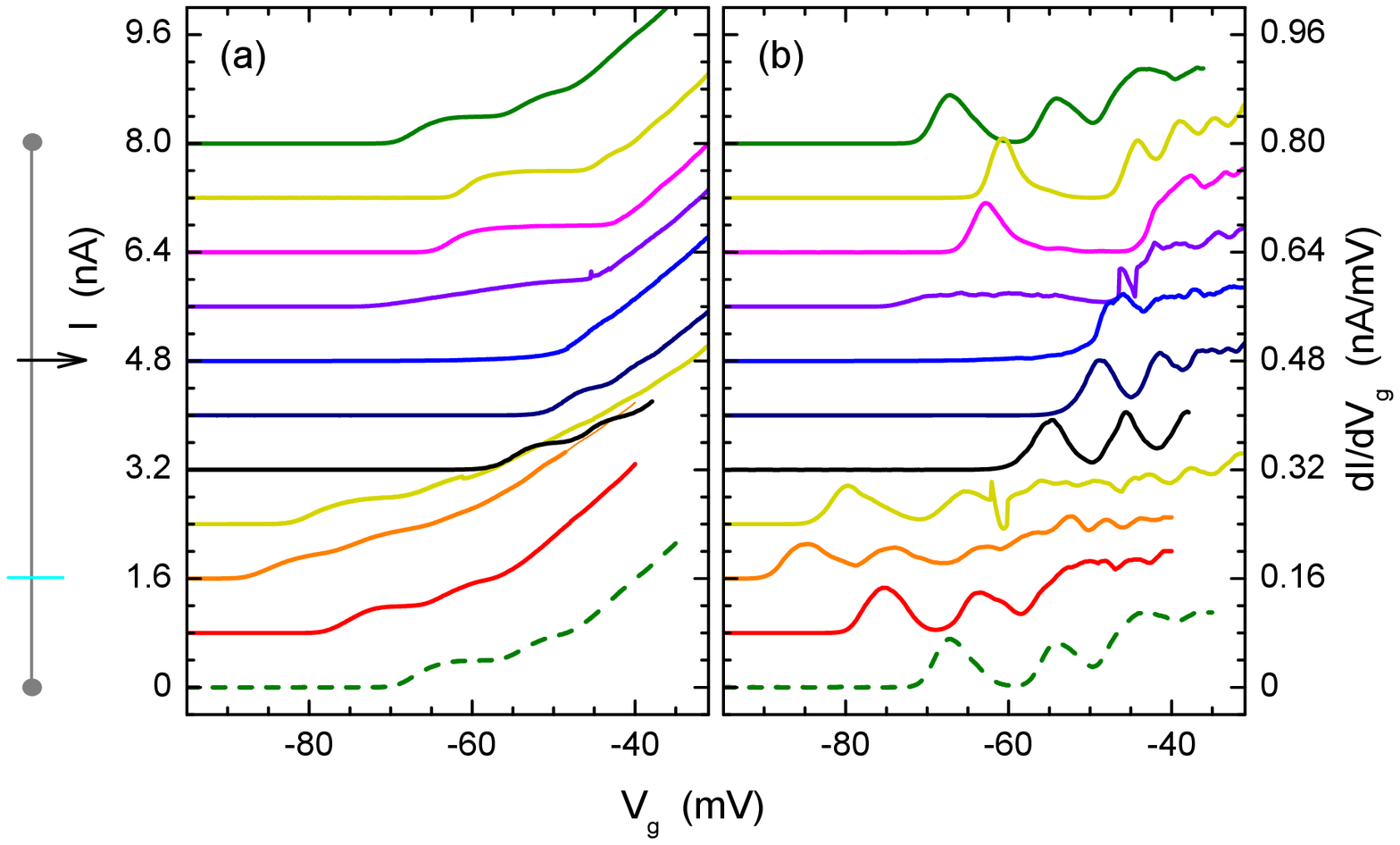}
 \end{center}
\caption{(Color on-line) (a) Acoustoelectric current $I$ and (b)
transconductance $dI/dV_g$ as a function of gate voltage $V_g$. The
consecutive traces were measured at different phase shifts between
the main and the counter-propagating SAW beam, in steps of about
$18^{\circ}$ (from top to bottom). The dashed curve completes the
$2\pi$ cycle. The traces are vertically offset for clarity. The
power applied to the main IDT was $P = +13.5\,$dBm, while the power
at the second IDT was attenuated by 8\,dB. Device 2E at $f =
2459.4\,$MHz and $T = 1.2\,$K.} \label{phase}
\end{figure}

Figure~\ref{results-2E} shows the selected $I(V_g)$ and
$dI(V_g)/dV_g$ traces from Figs.~\ref{3dim-2E} and
~\ref{greyscale-2E}, respectively. Consecutive curves were taken at
incremented frequencies, in steps of 0.1\,MHz. Note that the
frequency response of those curves is very similar to the one
described in the previous section:

(\emph{i}) When the frequency is increased from $f = 2458.6\,$MHz,
the onset of the acoustoelectric current shifts towards lower gate
voltages. At the same time, the AE plateaus broaden with respect to
$V_g$. This trend continues until $f = 2458.8\,$MHz.

(\emph{ii}) Around $f = 2458.8\,$MHz, a new set of densely-spaced
plateaus develops near the onset of $I(V_g)$. This new set evolves
smoothly until $f \approx 2459.6\,$MHz. Close to this frequency, the
current onset abruptly shifts towards larger gate voltages.

(\emph{iii}) The abrupt shift of the current onset marks the
transition to the next set of the AE plateaus, which corresponds to
the one in (\emph{i}).

As for device 2C, for some SAW frequencies, quantized steps in the
acoustoelectric current can be observed both at the expected
multiples of $I = nef$ and away from those values, see for example
the traces obtained at $f = 2459.9$ or 2460.0\,MHz.

The frequency response of the $I(V_g)$ characteristics
(Figs.~\ref{results-2C} and \ref{results-2E}) closely resembles
their response to the relative phase between the two
counter-propagating SAW beams (Fig.~\ref{phase}): When the frequency
(phase) is varied, two intervals can be distinguished within a
period of $1.1\,$MHz ($2\pi$) where two different sets of AE
plateaus dominate the $I(V_g)$ characteristics. Around certain
frequencies (phases), those two sets of quantized steps seem to
replace each other. This might result in abrupt shifts of the
current onset in $I(V_g)$, as for the traces indicated by the arrows
in Figs.~\ref{results-2C} and \ref{results-2E} (Fig. \ref{phase} for
phase). The other, more subtle, type of transition takes place for
the curves marked with the bars.

\begin{figure}
\includegraphics[width=1\textwidth]{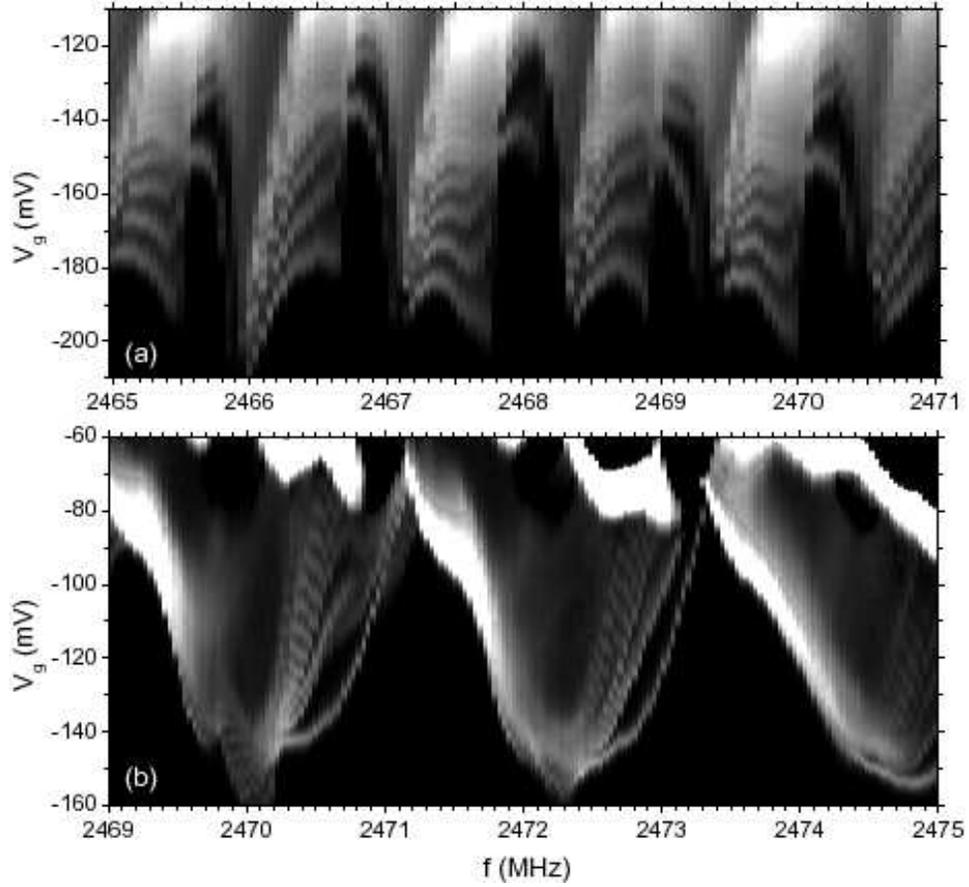}
\caption{Transconductance $dI/dV_g$ with respect to SAW frequency
$f$ and gate voltage $V_g$ after two different cool downs from room
temperature. Note the change in the period of frequency modulation
from (a) 1.1\,MHz to (b) 2.2\,MHz. The microwave power applied to
the SAW transducer differed slightly, being (a) +9.8 dBm and (b)
+10.8 dBm. Device 2C at $T = 1.8\,$K.} \label{double-beat}
\end{figure}

%
%
\subsection{Other beat periods}

The frequency response characteristics of our devices, similar to
those in Figs.~\ref{3dim-2C} - \ref{results-2E}, were stable as long
as the device was kept cold. When the sample was thermally cycled to
room temperature and cooled down again, the results differed in
details from those obtained previously. However, in most cases they
remained qualitatively similar to those described above.

Only on few cool-downs, a larger beat period than $1.1\,$MHz was
observed in the $I(V_g,f)$ characteristics.
Figure~\ref{double-beat}(b) shows an example where the dominant beat
period has changed to about 2.2\,MHz. This might suggest an enhanced
role of cross-talk, the interference between the dynamic SAW
potential and the airborne electromagnetic wave irradiated by the
transducer. The structure of the AE transitions in
Fig.~\ref{double-beat} is indeed very complex. At present, we cannot
provide a generalized description of those results like for the
1.1\,MHz oscillations in the preceding sections. Nevertheless, the
data in Fig.~\ref{double-beat} also show different sets of current
plateaus.

\section{DISCUSSION}

The frequency response measurements presented here reveal a
complicated pattern of the AE transitions. Their most apparent
feature, the 1.1\,MHz beating of the acoustoelectric current,
immediately reminds us of the interference patterns due to a
standing wave. In fact, Fig.~\ref{phase} demonstrates that the
results obtained while varying the SAW frequency closely resemble
those when the phase is controlled between two counter-propagating
SAW beams. The period of 1.1\,MHz corresponds to a $2\pi$ phase
shift. This already indicates that the standing wave matters, but it
does not specify where to find its nodes.

We have further demonstrated that two sub-intervals can be
distinguished within this period where two different sets of AE
plateaus dominate the $I(V_g)$ characteristics. At certain
frequencies, one set of quantized steps is replaced by the other
one. In some cases, they can even appear simultaneously in the
$I(V_g)$ traces, as if they were superposed on each other. In such a
situation, quantized steps can be resolved in the AE current both
near the expected multiples of $ef$ and at lower values, as shown in
Fig.~\ref{plateaus-2C}. Note that those two sets of AE plateaus
respond differently to the SAW frequency.

\subsection{Parallel channels?}

Different kinds of AE plateaus could be due to the presence of
parallel channels for the SAW-driven electrons. This could be
explained, for example, by branching of the electron flux at the
entrance to the QPC. Topinka {\it et
al.}\cite{Topinka2000,Topinka2001} recently demonstrated spatially
resolved images of the electron flow in the 2DEG, indicating
different current strands at the entrance and the exit of the point
contact. This was attributed to focusing of the electron
trajectories by a large ensemble of small ($\sim 0.1 \epsilon_F$)
ripples in the background potential, caused by impurities or donors.
In our experiments, such branching of the electron flux could affect
how the moving quantum dots are populated with electrons. On the
other hand, separate SAW channels could also be present in the QPC
itself. In the vicinity of a closed point contact, screening of
impurity potentials is reduced. Hence, the size of such potential
fluctuations could be much larger in the QPC region ($\sim 50\,$mV)
than in the 2DEG ($\sim 2\,$mV).\cite{Davies1989,Nixon1990} This, in
turn, could lead to the formation of separate pathways through the
constriction.

A quantized acoustoelectric current could be obtained separately for
each of those discrete trajectories. Note that changing the SAW
frequency (or phase) shifts the nodes and antinodes of the standing
wave. Therefore, the amplitude of the SAW modulation is affected at
a particular position within the QPC. At certain frequencies
(phases), different SAW channels could thus be opened or suppressed,
leading to abrupt transitions between different sets of quantized
steps in the acoustoelectric current. However, such a scheme should
be strongly dependent on the exact configuration of random impurity
potentials, whereas our results show systematic variations of the AE
plateaus. Therefore, we consider such a scenario rather unlikely.


\subsection{Effect of the short barrier(s)}

The SAW-driven single-electron transport is usually described using
a model of moving quantum dots, first suggested by Shilton {\it et
al.}.\cite{Shilton1996,Talyanskii1997} In such a model, single
electrons trapped in the local minima of the dynamic SAW potential
are transferred across the QPC barrier, which is assumed to be long
with respect to the SAW wavelength. As the dot moves towards the
center of the constriction and its size decreases, the Coulomb
repulsion between the trapped electrons restricts their number
inside the dot, forcing some of them to \emph{escape} back to the
2DEG reservoir they where captured from. Thus, the minimum size of
the dot determines the final number $n$ of transferred electrons. In
later theoretical studies,\cite{Aizin1998,Robinson2001} the role of
the electron \emph{escape} process for the AE current quantization
has been investigated in more detail. Another possible quantization
mechanism, the electron \emph{capture} process at the entrance to
the constriction, has been considered by Flensberg {\it et
al.}.\cite{Flensberg1999}

In all those models,\cite{Shilton1996,Talyanskii1997,%
Aizin1998,Flensberg1999,Robinson2001,Kashcheyevs2004} the static
barrier of the QPC is assumed to be long with respect to the SAW
wavelength. However, we believe that this is not entirely valid for
our devices. In spite of large nominal lengths of our QPCs
($\sim2\,\mu$m), their properties are determined by rather short
barriers of around $0.2\,\mu$m,\cite{Gloos2006} that is about $1/5$
of the SAW wavelength. Such a short barrier within the QPC channel
could reduce the number of electrons that are further carried in a
moving quantum dot.

We emphasize the main difference with respect to the models relying
solely on long QPC barriers\cite{Shilton1996,Talyanskii1997,%
Aizin1998,Flensberg1999,Robinson2001,Kashcheyevs2004} where only the
low-energy electrons at the bottom of the dot are transferred across
the constriction. This is no longer the case if the moving quantum
dot approaches a short (though high) barrier. Electrons in the
lowest energy states of the SAW minimum are then held back at the
barrier and return to the reservoir they originated from. Only those
with higher energy are transferred across the QPC and contribute to
the AE current.

Note that expelling of low-energy electrons at the short static
barrier can take place in series with the quantization mechanisms
described in previous models, relying on the escape or capture of
high-energy electrons.\cite{Shilton1996,Talyanskii1997,Aizin1998,%
Flensberg1999,Robinson2001,Kashcheyevs2004} The originally quantized
current can thus be further reduced at the short barrier, also to a
new quantized level. This could explain Fig.~\ref{plateaus-2C}: The
higher order plateaus (thin lines/points) are lowered when the
second mechanism (thick line/points) sets in. Thus, this simple
model would provide an explanation to the presence of at least two
different kinds of AE current plateaus. It could also explain the
transition between the series of broad and well-pronounced AE
plateaus starting at low $V_g$, and the narrowly-spaced and
barely-pronounced quantized steps at larger gate voltages, as
observed in the $I(V_g)$ characteristics in Figs.~\ref{typical},
\ref{results-2C}, and \ref{results-2E}.

One could further speculate that, at very narrow channels, more than
one barrier is present in the constriction. One set of the AE
plateaus could then result from operating the device as a static
quantum dot with two separate barriers of comparable magnitude, as
found by Fletcher {\it et al.}\cite{Fletcher2003} Whereas a second
set could be due to a single barrier. In few cases for some of our
devices, we indeed observed that the Coulomb blockade peaks at zero
SAW power, indicating the formation of a static quantum dot, evolved
into the AE current plateaus when the SAW amplitude was
increased.\cite{Utko2005}


\section{CONCLUSION}

The quantization of the AE current as a function of the gate voltage
and the SAW frequency reveals a rather involved picture with several
transitions between distinct transport regimes. Changing the SAW
frequency has the same effect as changing the phase of a
counter-propagating SAW beam. This indicates the presence of a
standing wave, which either enhances or reduces locally the
potential profile along the QPC channel. This profile determines the
quantization condition for the AE current. Still, the exact position
of the standing-wave node, and whether it reduces or increases the
electrostatic potential there, remains unknown.

We have distinguished two different regimes of the AE current flow.
A discrete number of electrons can be captured and transported
across the QPC in the minima of the SAW potential, as already
described in
Refs.~\onlinecite{Shilton1996,Talyanskii1997,Aizin1998,%
Flensberg1999,Robinson2001,Kashcheyevs2004}. The second mechanism
results from the short barriers of our QPCs. Their
presence\cite{Gloos2006} already implies that the conventional
picture of a conveyor-belt-like transport across the QPC might not
be entirely valid for our devices. The number of electrons
transferred per SAW cycle across the short barrier is determined by
its height: Only high-energy electrons can pass it in forward
direction. Thus, it is possible that an already quantized current
can be further changed (reduced) to a new quantized condition.

\section*{ACKNOWLEDGMENTS}
This work was supported by the European Commission FET Project
SAWPHOTON. P.U. acknowledges support from EC FP6 funding (contract
no. FP6-2004-IST-003673).

\end{document}